\documentclass[pra,aps,twocolumn,epsfig,superscriptaddress,showpacs]{revtex4}
\usepackage{amssymb}
\usepackage{amsmath}
\usepackage{epsfig}
\usepackage{amsfonts}
\usepackage{color}
\usepackage{float}
\usepackage{latexsym}
\usepackage{mathrsfs}
\usepackage{feynmf}
\def\QED{\leavevmode\unskip\penalty9999 \hbox{}\nobreak\hfill
     \quad\hbox{\leavevmode  \hbox to.77778em{%
               \hfil\vrule   \vbox to.675em%
               {\hrule width.6em\vfil\hrule}\vrule\hfil}}
     \par\vskip3pt}
\def\qed{\leavevmode\unskip\penalty9999 \hbox{}\nobreak\hfill
     \quad\hbox{\leavevmode  \hbox to.77778em{%
               \hfil\vrule   \vbox to.675em%
               {\hrule width.6em\vfil\hrule}\vrule\hfil}}
\par\vskip3pt}
\def\ibb #1{\leavevmode\hbox{\kern.3em\vrule
     height 1.5ex depth -.1ex width .4pt\kern-.3em\rm#1}}

\newcommand{\be}{\begin{equation}}
\newcommand{\ee}{\end{equation}}
\newcommand{\ba}{\begin{array}}
\newcommand{\ea}{\end{array}}
\newcommand{\bqa}{\begin{eqnarray}}
\newcommand{\eqa}{\end{eqnarray}}

\newcommand{\bra}[1]{\ensuremath{\langle #1 |}}
\newcommand{\ket}[1]{\ensuremath{| #1 \rangle}}

\newcommand{\beq}{\begin{eqnarray}}
\newcommand{\eeq}{\end{eqnarray}}

\begin{document}

\author {Zhi-Hao Ma$^{1}$, Zhi-Hua Chen$^{2}$, Jing-Ling Chen$^{3}$}

\affiliation { Department of Mathematics, Shanghai
Jiaotong University, Shanghai, 200240, P. R. China }

\affiliation {Department of Science, Zhijiang college, Zhejiang
University of technology, Hangzhou, 310024, P.R.China}

\affiliation {Theoretical Physics
Division, Chern Institute of Mathematics, Nankai University,
Tianjin, 300071, P.R.China}

\date{\today}

\begin{abstract}

we derive criteria for $k$-separability of multipartite Quantum state
\end{abstract}

\title{Criteria For $K$-Separability Of Multipartite Quantum State}
 \maketitle

\section{Introduction}

Entanglement, a central feature
of quantum mechanics, lies at the heart of quantum information theory, and was found many applications in quantum information processing tasks
such as quantum cryptography,  quantum teleportation, and
measurement-based quantum computing.

In any of the above experiments, a natural question arise: How can we be sure that entanglement exists? For two-partite quantum states, many results were obtained, and is being more and more understood~\cite{horodeckiqe}.
However, for multi-partite state, the situation becomes troublesome. From experiment point of view, genuine multipartite entanglement is most important(e.g. ~\cite{horodeckiqe}~\cite{bio}), but its detection only beginning to be developed(see e.g. Refs. ~\cite{2002Seevinck,2002Collins,Devi07,Gillet08,Hassan09,guehnewit, guehnecrit, hmgh, brukner,Chen2011,Gisin2011}). So it is emergent to find new method to detect  genuine  Multipartite entanglement.

In this brief report, we aim to derive a criteria for $k$-separability of quantum states.

{\bf Definition 1:} An $n$-partite pure quantum state $\left|\Psi_{k-sep}\right\rangle$ is called $k$-separable, iff it can be written as a product of $k$ substates:
\beq \left|\Psi_{k-sep}\right\rangle = \ket{\Psi_1}\otimes\ket{\Psi_2}\otimes \cdots \otimes \ket{\Psi_k} \eeq
A mixed state $\rho_{k-sep}$ is called $k$-separable, iff it has a decomposition into $k$-separable pure states:
\beq \rho_{k-sep} = \sum_{i} p_i \ket{\Psi_{k-sep}^i}\bra{\Psi_{k-sep}^i} \eeq
In particular, an $n$-partite state is called fully separable, iff it is $n$-separable. It is  called genuinely $n$-partite entangled, iff it is not biseparable (2-separable). Note that the individual pure states composing a $k$-separable mixed state may be $k$-separable under different partitions. Hence, in general, $k$-separable mixed states are not separable w.r.t. any specific partition, which makes $k$-separability rather difficult to detect. Let us also remark that whenever a state is $k$-separable, it is automatically also $k'$-separable for all $k' < k$
\\
{\bf Definition 2.} In order to formulate our criterion for $k$-separability, we first need to define permutation operators $P_{ij}$ acting on two copies of an $n$-partite state. These operators swap the $i$-th and $j$-th subsystems of the two copies respectively:
\[\begin{array}{lll}  P_{ij} \ket{\Psi_{a_1,a_2,\cdots,a_n}}\otimes\ket{\Psi_{b_1,b_2,\cdots,b_n}}&\\
 =\ket{\Psi_{a_1,a_2,\cdots,a_{i-1},b_j,a_{i+1},\cdots,a_n}}\otimes\ket{\Psi_{b_1,b_2,\cdots,b_{j-1},a_i,b_{j+1},\cdots,b_n}} &\\
\end{array}\]
where the $a_i$ and $b_j$ indicate the subsystems of the first and second copy of the state, respectively.\\

\section{Criterion for $k$-separability}

We can get a  Criterion for $k$-separability as following:
\emph{Theorem:} Every $k$-separable state $\rho$ satisfies
\begin{equation} \label{}
\sqrt{\bra{\Phi}\rho^{\otimes 2}P_{tot}\ket{\Phi}}-\sum_{\{\alpha\}}\left(\prod_{i,j=1}^{k}\bra{\Phi}P^{\dagger}_{\alpha_{ij}}\rho^{\otimes 2}P_{\alpha_{ij}}\ket{\Phi}\right)^{\frac{1}{2k^{2}}}\leq 0
\end{equation}
for all fully separable states $\left|\Phi\right\rangle$, where the sum runs over all possible partitions $\alpha$ of the considered system into $k$ subsystems, the permutation operators $P_{\alpha_{ij}}$ are the operators permuting the two copies of all subsystems contained in the $i$-th subset and $j$-th subset of the partition $\alpha$ respectively and $P_{tot}$ is the total permutation operator, permuting the two copies.\\
\\

\emph{Proof:} To prove this, observe that (like in Refs.~\cite{guehnecrit,hmgh}) the the inequality is a convex function of $\rho$ (since the first term is the absolute value of a density matrix element and each term in the sum is the $2k$-th root of the product of $2k$ density matrix diagonal elements). Consequently, it suffices to prove the validity for pure states and validity for mixed states is guaranteed. So, let us assume w.l.o.g., that the given pure state $\rho$ is $k$-separable w.r.t. the $k$-partition $\tilde{\alpha}$. Due to its separability, $\rho$ is invariant under permutation of each element of $\tilde{\alpha}$:
\beq P_{\tilde{\alpha}_{ij}}^\dagger \rho^{\otimes 2} P_{\tilde{\alpha}_{ij}} = \rho^{\otimes 2} . \eeq
Therefore, the corresponding term in the sum can be written as
\[\begin{array}{lll}\left(\prod_{i,j=1}^{k} \bra{\Phi} P_{\tilde{\alpha}_{ij}}^\dagger \rho^{\otimes 2} P_{\tilde{\alpha}_{ij}} \ket{\Phi}\right)^{\frac{1}{2k^{2}}} =
\left(\prod_{i,j=1}^{k} \bra{\Phi} \rho^{\otimes 2} \ket{\Phi}\right)^{\frac{1}{2k^{2}}} &\\
= \left(\prod_{i,j=1}^{k} \sqrt{\bra{\phi_1}\rho\ket{\phi_1}\bra{\phi_2}\rho\ket{\phi_2}} \right)^{\frac{1}{k^{2}}}= \sqrt{\bra{\phi_1}\rho\ket{\phi_1}\bra{\phi_2}\rho\ket{\phi_2}}&\\
\end{array}\]

where we used $\ket{\Phi}=\ket{\phi_1}\otimes\ket{\phi_2}$. Using $P_{tot} \ket{\phi_1}\otimes\ket{\phi_2} = \ket{\phi_2}\otimes\ket{\phi_1}$, we can now rewrite ineq. (4) as

\[\begin{array}{lll} \left|\bra{\phi_1} \rho \ket{\phi_2}\right| - \sqrt{\bra{\phi_1}\rho\ket{\phi_1}\bra{\phi_2}\rho\ket{\phi_2}}&\\
- \sum_{\{\alpha\neq\tilde{\alpha}\}}\left(\prod_{i,j=1}^{k} \bra{\Phi} P_{\alpha_{ij}}^\dagger \rho^{\otimes 2} P_{\alpha_{ij}} \ket{\Phi}\right)^{\frac{1}{2k^{2}}} \leq 0 &\\
\end{array}\]

Now, the first term is an off-diagonal matrix-element and the second term is the squareroot of the product of the two corresponding diagonal elements, hence (and because $\rho$ is a pure state), the first two terms cancel each other. It is evident that the remaining sum over strictly nonnegative terms (diagonal elements) with a negative sign is non positive, which proves our theorem.\qed

\vskip 0.1 in {\noindent\bf Acknowledgment.} This work is supported
by NSF of China(10901103), partially supported by a grant of science
and technology commission of Shanghai Municipality (STCSM, No.
09XD1402500).

\end{document}